\documentstyle[11pt,psfig,aaspp4]{article}

\def\kms{km~s$^{-1}$~}

\def\W50{W$_{50}$~}
\received{7 March 1996}
\accepted{Someday 1999}
\journalid{337}{}
\articleid{11}{14}

\begin{document}
\title{A Test of the Lauer--Postman Bulk Flow}

\author {Riccardo Giovanelli, Martha P. Haynes}
\affil{Center for Radiophysics and Space Research
and National Astronomy and Ionosphere Center,
Cornell University, Ithaca, NY 14953}

\author {Gary Wegner}
\affil{Dept. of Physics and Astronomy, Dartmouth College, Hanover, 
NH 03755}

\author {Luiz N. da Costa}
\affil{European Southern Observatory, Karl Schwarzschild Str. 2, D--85748
Garching b. M\"unchen, Germany and Observatorio Nacional, Rio de Janeiro, Brazil}

\author {Wolfram Freudling}
\affil{European Southern Observatory and Space Telescope--European Coordinating Facility, Karl 
Schwarzschild Str. 2, D--85748
Garching b. M\"unchen, Germany}

\author {John J. Salzer}
\affil{Dept. of Astronomy, Wesleyan University, Middletown, CT 06457}


\begin{abstract}

We use Tully--Fisher distances for a sample of field late spiral galaxies to 
test the Lauer and Postman (1994; hereinafter LP) result suggestive of a bulk 
flow with respect to the Cosmic Microwave Background reference frame, of 
amplitude of +689 \kms in the direction $l=343^\circ$, $b=+52^\circ$. A total 
of 432 galaxies are used, subdivided between two cones, of 30$^\circ$ semiaperture 
each and pointed respectively toward the apex and antapex of the LP motion.
The peculiar velocities in the two data sets are inconsistent 
with a bulk flow of the amplitude claimed by LP. When combined in opposition,
the peculiar velocity medians in shells of constant redshift width are never 
larger than half the amplitude of the LP bulk flow. Out to 5000 \kms 
the median bulk velocity in the LP apex--antapex cones is about 200 \kms
or less, dropping to a value indistinguishable from zero beyond that distance.
It can be excluded that field spiral galaxies within 8000 \kms partake of a
bulk flow of the amplitude and direction reported by LP.

\end{abstract}


\section {Introduction}

The dipole moment of the sky brightness distribution in the cosmic microwave background
radiation (CMB) is thought to arise from the Doppler shift due to the deviation in the 
motion of the Local Group (LG) from smooth Hubble flow. That motion appears to be directed
towards galactic coordinates $(276^\circ,+30^\circ)$, with amplitude $627\pm22$
\kms (Kogut et al. 1993). Inhomogeneities in the mass distribution will produce deviations
from smooth Hubble flow, also referred to as ``peculiar velocities''. Therefore the
CMB dipole raises the challenges of (a) identifying the sources of the LG motion and
(b) mapping the peculiar velocity field in the local universe, of which the LG motion
is a single point sample. The coherence length scales and characteristic amplitudes
of the peculiar velocity field are important cosmological parameters, which can in
principle be obtained via the systematic determination of distances {\it and} redshifts
of large samples of galaxies, which respond as point sources to the buffeting of
the large--scale gravitational field. Alternatively, the moments of the light distribution 
of all--sky catalogs of extragalactic objects can be used. These techniques have been
extensively applied, producing often conflicting results.
 
After the early measurements of Rubin et al. (1976) of the flux dipole of ScI galaxies, 
Aaronson et al. (1986 and refs. therein)
made headway in the measurement of the infall of the Local Group towards the Virgo
cluster. As Virgo is located nearly a radian off the CMB dipole apex, attention 
shifted towards the peculiar velocity field outside the Local Supercluster, with the 
expectation of finding a set of galaxies that would provide a reference frame with 
respect to which the amplitude and direction of the LG motion would match that revealed
by the CMB dipole. Several authors pointed out the important role that might be played 
by the Hydra--Centaurus supercluster (Shaya 1984, Sandage and Tammann 1985, Lilje et al. 
1986), until the existence of a ``Great Attractor'' (GA), a mass concentration largely 
obscured by Milky Way dust, was invoked by the work of Lynden--Bell et al. (1988) and 
Dressler et al. (1987); the GA was placed at a redshift of 4300 \kms, about 10\% farther
than Hydra. More recent work, however, has raised the possibility that the local universe 
may instead be partaking in a flow of larger scale, whereby the component of largest 
amplitude in the LG motion has a coherence length a few times larger than the distance 
to the GA. Scaramella et al. (1989) suggested that the Shapley Supercluster, at about 
13,000 \kms and in the general direction of the GA, may constitute an important density inhomogeneity in determining our motion; Willick (1990) reported that the Perseus--Pisces 
Supercluster, located at a distance of about 5000 \kms and roughly antipodal to the GA, 
moves towards us and thus towards the GA at about 400 \kms.  Courteau et al. (1993) 
reinforced that finding by reporting the measurement of a ``bulk'' flow, i.e. the global 
motion of all matter within a top hat average out to 6000 \kms, of $385\pm38$ \kms toward 
galactic coordinates $(294^\circ,0^\circ)$. The interested reader is addressed to
the recent comprehensive review of Strauss and Willick (1995).

Mathewson et al. (1992) (MFB) carried out an 
extensive survey of galaxy distances in the Southern hemisphere and reported lack of 
backflow in the GA, corroborating the view that Perseus--Pisces, LG and GA share 
a bulk motion, that may be produced by more distant gravitational sources; it was even 
suggested that the motion may have a non--gravitational origin (Mathewson 1995). In 1994,
an important result was contributed by Lauer and Postman: they measured the dipole of
the distribution of the brightest elliptical galaxies in 119 clusters, distributed over
a volume of 15,000 \kms radius and with an effective depth of approximately 10,000 \kms. 
They found that {\it the reference frame defined by the group of clusters is in motion with 
respect to the CMB, at a velocity of $689\pm178$ \kms, and towards the direction of galactic 
coordinates $(343^\circ, +52^\circ)(\pm23^\circ)$}. A reanalysis of the LP data
by Colless (1995) confirmed their result. Not only is this motion roughly orthogonal 
to the CMB apex and in large disagreement with that reported by other surveys; it also poses
an embarassment to many models that attempt to reproduce the characteristics of large scale 
structure, as they encounter difficulties in accomodating bulk flows of large amplitude 
and scale (e.g. Dekel 1994, Strauss et al. 1995). The work of Riess et al. (1995), 
which uses a sample
of 13 SN Ia (now extended to 20 according to a personal communication of R. Kirshner), 
disagrees with the LP bulk motion result, rather suggesting agreement with the CMB apex.

Here we report on a simple test of the LP result. We utilize a set of distances to field
('SFI') and cluster ('SCI') spiral galaxies, obtained using the Tully--Fisher (1977) 
technique in the I band. They are part of
an all--sky sample of more than 2,000 objects; while preliminary results of the 
survey have been reported by Giovanelli et al. (1994,1995), Freudling et al. (1995) and
da Costa et al. (1996a), a more general
and detailed analysis of these data will be presented elsewhere.

\section {Data Selection} 

We have obtained CCD I--band images and velocity widths for a sample of late spiral
galaxies north of declination $-40^\circ$, which we have combined with the 
southern data data published by MFB to obtain
an all sky sample of spiral galaxy distances extending to about 8000 \kms. Sample
selection criteria are described in Giovanelli et al. (1994). Detailed descriptions of 
data reduction and analysis are in preparation. 

A test of the LP result using this data set is relatively straightforward, thanks
to the high galactic latitude of the LP apex. Since our sample of field galaxies
ends in regions of high extinction and thus of low galactic latitude, the testing
of bulk flows with apices close to the galactic plane, such as that reported by 
Courteau et al. (1993), requires a more careful reconstruction of the peculiar
velocity field, which we will present elsewhere (da Costa et al. 1996b). Here we 
restrict ourselves to testing the LP result. 

We select two subsets of the data, corresponding to cones each subtending an 
angle of $30^\circ$ half--aperture, centered respectively on the apex 
[$(l,b)=(343^\circ,+52^\circ)$]
and antapex directions of the LP bulk flow. The antapex cone includes 235 galaxies with
measured distances, of which 84\% are drawn from our observations and 16\% from those
of MFB; the apex cone includes 197 galaxies, 2/3 from our own and 1/3 from MFB's 
observations. The optical rotation widths of MFB have been re--analyzed to obtain 
consistency with the 
techniques applied to our data. MFB contain a mixture of 21cm radio 
widths and widths derived from optical rotation curves. An important
concern has been that of removing the bias introduced by optical rotation curves
which insufficiently sample the galactic disks, and thus appear to be still rising
at the last measured point; these objects, which tend to be more frequent in the GA 
region, sometimes yield severe underestimates of
the velocity widths, and, as a consequence, systematically large positive peculiar
velocities. Several objects have been rejected for this reason, while for others
velocity widths at a radius encompassing 83\% of the light (``optical radius'')
were estimated, using
the folded versions of the rotation curves of MFB produced by Persic and
Salucci (1995). In cases in which the MFB rotation curve did not reach as far
out as the optical radius as defined above, an extrapolation was carried out,
using the Persic and Salucci (1991) ``universal rotation curve'' parametrization,
which relates the shape of a rotation curve with the galaxy's luminosity.
A total of 69 of the MFB galaxies used here have velocity widths derived from
optical rotation curves; the derived widths at the optical radius appear
to be reliable estimates. It should be pointed out that the remainder of MFB's
widths as well as all of our own were obtained at 21cm, and therefore are
unaffected by recalibration. The recalibrated MFB data follow the same TF
relation as those with 21cm widths.

\begin{figure} 
\centerline{\psfig{figure=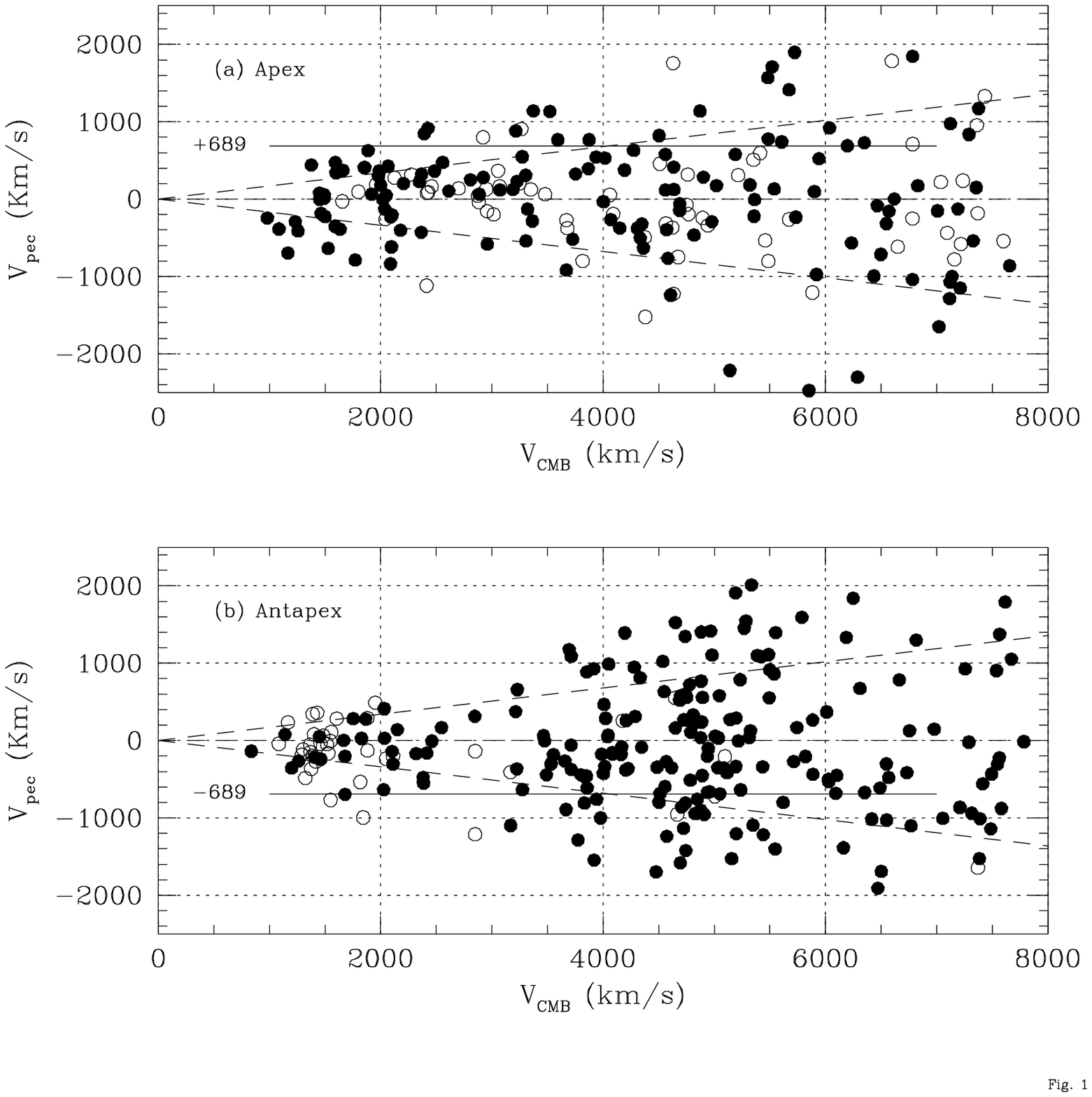}}
\caption {(a) Peculiar velocity of 
spiral galaxies in the LP apex cone; the abscissa is the radial velocity in the CMB 
reference frame. Dashed lines identify a mean scatter in the TF relation of 0.35 mag. 
The +689 \kms level corresponds to the amplitude of the LP bulk velocity, inset for 
comparison. Filled symbols refer to galaxies observed by us, while unfilled symbols
correspond to data from MFB. (b) Analogous plot to that 
in panel (a), except that it refers to the LP antapex cone. A small number of objects
lie outside the plot margins.}
\end{figure}

\section {Motions in the LP Apex and Antapex Cones} 

Peculiar velocities are computed for each galaxy from the observed radial velocity
measured in the CMB reference frame, $v_{cmb}$, and the Hubble velocity obtained 
from the inverse TF relation, $v_{tf}$, via $v_{pec} = v_{cmb} - v_{tf}$. The
inverse relation was used because, as shown by Freudling et al. (1995), the bias
that affects the estimates of $v_{pec}$ is significantly reduced when the
inverse rather than the direct TF relation is adopted. Peculiar velocities are
not corrected for the inhomogeneous bias; this correction is not necessary for
our purposes, as we discuss when we present fig. 2 further below.

In fig. 1, we display the peculiar velocities of the galaxies in the LP apex and 
antapex cones, plotted versus the galaxy radial velocity, measured in the CMB reference 
frame. Rather than plotting error bars for each point, we inset dashed lines which reflect 
the mean error on the peculiar velocity, as derived from 
a mean scatter in the Tully--Fisher relation of 0.35 mag. The amplitude of the LP 
motion is also inset as a solid horizontal line at $\pm689$ \kms . Filled symbols
identify objects observed by us, while unfilled ones are from MFB.
Note that the horizontal axis in fig. 1 is the galaxy redshift $v_{cmb}$ rather than 
the Tully--Fisher velocity $v_{tf}$, and therefore the reader is cautioned
against directly reading of the horizontal axis as a distance parameter.
 
Galaxies of four cluster regions are part of the apex and antapex samples; in fig. 1,
they are plotted at their own $v_{cmb}$ rather than that of the cluster. Galaxies 
in the clusters ESO 508 and A3574 are in the apex region: 
ESO508 is near $V_{cmb} = 3200$ \kms and A3574 near $V_{cmb} = 4800$ \kms. Galaxies 
in the cluster A400 and in the Eridanus group, respectively near 
$V_{cmb} = 7000$ \kms and near 1500 \kms, are in the antapex region. 
The systemic peculiar velocities of the four clusters are $-128\pm235$ \kms for A400,
$-343\pm70$ \kms for Eridanus, $+465\pm140$ \kms for ESO508 and $+44\pm185$ \kms 
for A3574 (also known as Klemola 27). The large amplitude in the peculiar velocity
field in the antapex region between 4000 and 5500 \kms is associated with infall
and backflow in the
high density regions of the Perseus--Pisces supercluster.

The $30^\circ$ semi--aperture of the apex and antapex cones implies that if all 
galaxies in the sample shared a bulk velocity directed along the LP dipole, the 
radial peculiar velocity component of any galaxy should be at least 87\% the bulk 
velocity amplitude. 
The plain visual inspection of fig. 1 reveals that, over the redshift regime sampled
by our data, the mean motion of the spiral galaxies falls well short of that required
by the LP finding. The scatter in the data is however large. It should also be
pointed out that the LG motion of 630 \kms with respect to the CMB, if shared by 
galaxies in the sample, should project at roughly half that amplitude or about 300 \kms, 
in the direction of the LP apex and antapex.

An alternative representation of the comparison between our data and the LP bulk flow 
can be obtained by combining the apex and antapex regions, as done in figure 2. We bin 
the data in redshift windows of 1000 \kms (except for the last, that includes objects
between 6000 and 8000 \kms), combining in opposition the apex and
antapex data sets, and obtain for each bin a median value of the distribution of
peculiar velocities. The thin error bars about each point subtend the range within which 
the second and third quartile of the distribution of peculiar velocities are enclosed. 
Mean values and standard deviations are not used because
the distributions depart substantially from a normal one, partly due to real motions
and in part due to the presence of a few very high peculiar velocity points (which are 
unlikely to have a physical origin, and may reflect undetected inadequacies of the data 
for Tully--Fisher application).
The number of galaxies contributing to each data point are given in the lower part
of the graph. A rough estimate of the accuracy of the estimate of the median values 
is indicated by the thick portion of the error bars; this estimate is obtained 
by dividing half the peculiar velocity range, over which two thirds of the points are found,
by the square root of the number of points in each bin, producing a number that for a
normal distribution would be close to the standard error on the mean.

As we pointed out at the beginning of this section, the peculiar velocities are uncorrected
for Malmquist bias. As shown by our Monte Carlo simulations (Freudling et al. 1995),
if peculiar velocities are averaged in bins of redshift, rather than TF velocity, and
if the inverse TF relation is used to estimate $v_{pec}$, then no significant Malmquist 
bias is present in the averages (see their fig. 5d). Any small, residual Malmquist bias
present would, at any rate, be cancelled out by the combination of peculiar velocities
in the apex and antapex cones shown in fig. 2. These results are corroborated by
the estimate of the bulk motion from the reconstruction of the 3D velocity field
within a 6000 \kms top--hat, obtained by da Costa et al. (1996b). 

\begin{figure} 
\centerline{\psfig{figure=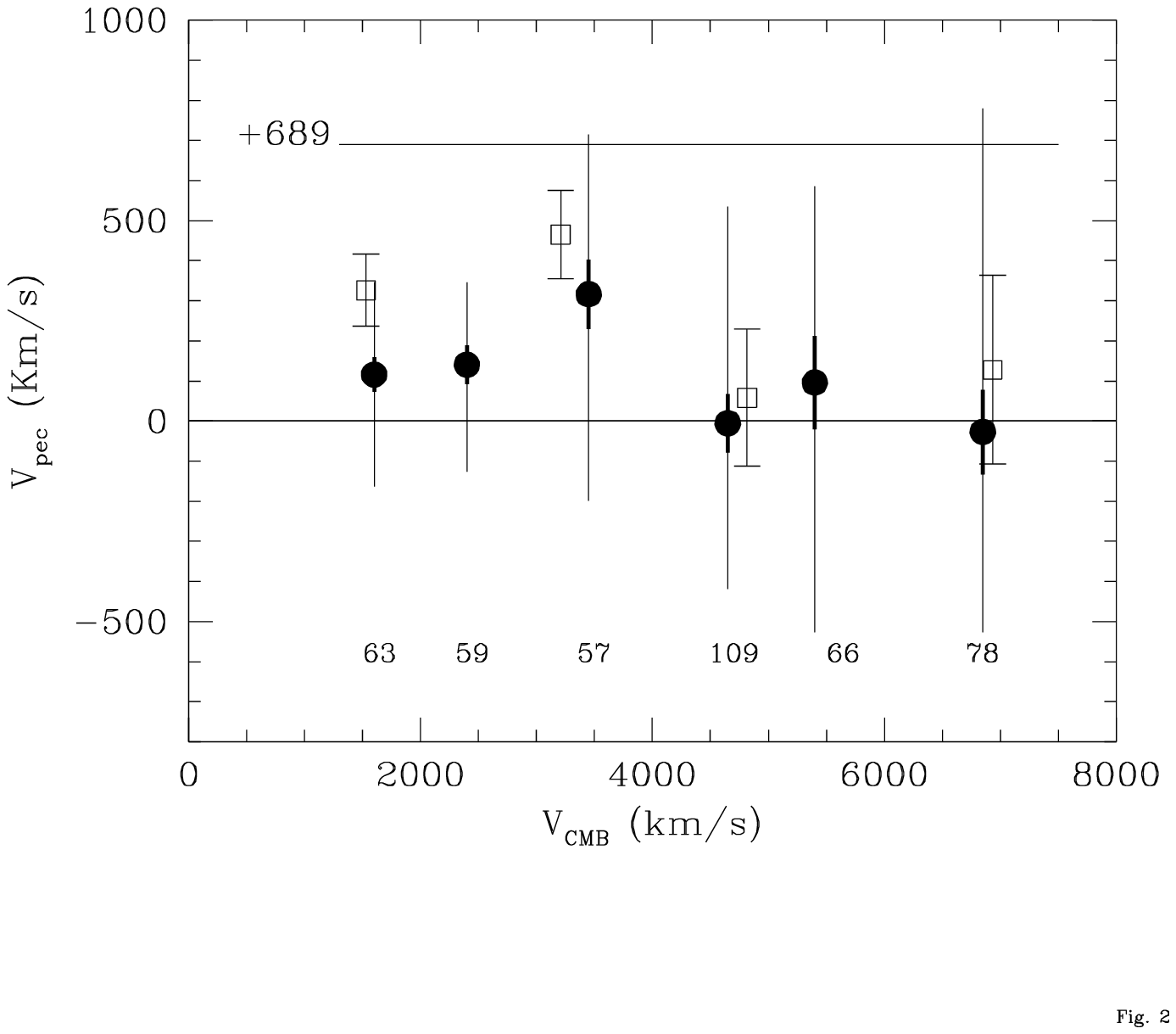}}
\caption{Median peculiar velocities, binned by shells of radial velocity 1000 \kms
wide; values refer to the total bulk velocity of the apex and antapex regions
in each shell, added in opposition so that the sign is that of the net velocity 
in the LP apex direction. Thin error bars refer to the range spanned by data within the
two inner quartiles, and thick error bars identify our best guess of the accuracy of 
the median value determination, akin to a standard error on the mean. Numbers under each symbol identify the number of galaxies used in each bin.
The LP bulk flow amplitude of 689 \kms is inset as a horizontal line for comparison. 
The peculiar velocities of the four clusters in the apex--antapex cone are plotted as unfilled symbols; they are, from low to high $V_{CMB}$, Eridanus, ESO508, A3574 and
A400. The plotted velocity of A400 and Eridanus are the opposite of the measured 
velocities, as a positive bulk velocity is one directed in the apex direction.
}
\end{figure}

Every median value in figure 2 lies below the LP bulk velocity amplitude; in fact,
most points do not exceed half that amount. The median value of the bulk velocity
within a volume of 5000 \kms redshift radius is less than 200 \kms, while that
in the volume between 4500 and 8000 \kms appears indistinguishable from zero. The component
of the LG motion with respect to the CMB (as obtained from the CMB dipole), along
the direction of the LP apex--antapex amounts to about 300 \kms. This compares 
reasonably well 
with the amplitude of the motions of the galaxies within 4000 \kms, and is
consistent with the idea that those may be in part travelling companions of the LG with 
respect to the CMB. Beyond that distance, the amplitude of the motion appears 
reduced; a more detailed interpretation is hampered by the presence of large infall 
and backflow motions in the Perseus--Pisces supercluster at 5000 \kms, which are 
visible in the LP antapex cone. In figure 2 we also plot as unfilled squares the 
peculiar velocities of the four clusters which lie within the apex--antapex cone, 
namely Eridanus, ESO508, A3574 and A400, as derived by Giovanelli et al. (1996); 
for A400 and Eridanus, which are in the antapex region, the plotted values are the
opposites of the measured velocities, as a positive bulk velocity is one directed 
in the apex direction. The motions of the clusters agree, within the accuracy of the 
measurements, with the individual galaxies' medians.

\section {Summary} 

We have analyzed the motion of spiral galaxies in two cones, each 
of $30^\circ$ semi--aperture, centered respectively about the apex and the antapex 
directions of the bulk flow reported by Lauer and Postman (1994).
The median amplitudes of the bulk velocities identifiable in our data, in any redshift
shell within 8000 \kms, are generally less than half the LP result. The median peculiar
velocity within a redshift of 5000 \kms hovers between 0 and 350 \kms, depending on
the redshift bin in which it is measured, and the flow
appears to subside at redshifts of about 4500 \kms and larger.  
An average bulk flow with respect to the CMB as large as 689 \kms, of galaxies 
within the volume subtended by these data, which extend out to 8000 \kms,  
can be excluded. This result is corroborated by the analysis presented by
da Costa et al. (1996b). What cannot be excluded, of course, if that much of the LP
flow signature occurs in shells external to the maximum redshift sampled
by this data: we would then be presented with the unusual geometry of
an envelope in coherent, large bulk flow, about a core region which would
not share in that coherence.

\vfill
\acknowledgements

The results presented in this paper are based on observations carried out at
the National Astronomy and Ionosphere Center (NAIC), the National Radio Astronomy
Observatory (NRAO), the Kitt Peak National Observatory (KPNO), the Cerro Tololo 
Interamerican Observatory (CTIO), the Palomar Observatory (PO), the Observatory of 
Paris at Nan\c cay, the Michigan--Dartmouth--MIT Observatory (MDM) and the European
Southern Observatory (ESO). NAIC, NRAO, KPNO 
and CTIO are respectively operated by Cornell University, Associated Universities, inc., 
and Associated Universities for Research in Astronomy, all under cooperative agreements
with the National Science Foundation. Access to the 5m telescope at PO was guaranteed 
under an agreement between Cornell University and the California Institute of Technology. 
This research was supported by NSF grants AST94--20505 to RG, AST90-14850 and AST90-23450 
to MH and AST93--47714 to GW.

\end{document}